\documentclass[12pt]{article}
\begin{document}
\setcounter{page}{1}
\def\theequation{\arabic{section}.\arabic{equation}}
\def\theequation{\thesection.\arabic{equation}}
\setcounter{section}{0}

\title{On the hadron production\\ from the quark--gluon plasma phase
in ultra--relativistic heavy--ion collisions}

\author{A. Ya. Berdnikov, Ya. A. Berdnikov~\thanks{E--mail:
berdnikov@twonet.stu.neva.ru, State Technical University, Department
of Nuclear Physics, 195251 St. Petersburg, Russian Federation} ,\\
A. N. Ivanov~\thanks{E--mail: ivanov@kph.tuwien.ac.at, Tel.:
+43--1--58801--14261, Fax: +43--1--58801--14299}~$^\|$ , 
V. A. Ivanova~\thanks{Participation within the Scientific 
Student Program of the Department of Nuclear Physics in  
State Technical University of St. Petersburg, Russian Federation} ,
V. F. Kosmach~\thanks{State Technical University, Department of
Nuclear Physics, 195251 St. Petersburg, Russian Federation} ,\\
V. M. Samsonov~\thanks{E--mail: samsonov@hep486.pnpi.spb.ru,
St. Petersburg Institute for Nuclear Research, Gatchina, Russian
Federation} ,  N. I. Troitskaya~\thanks{Permanent Address: State
Technical University, Department of Nuclear Physics, 195251
St. Petersburg, Russian Federation}}

\date{\today}

\maketitle

\begin{center}
{\it Institut f\"ur Kernphysik, Technische Universit\"at Wien, \\
Wiedner Hauptstr. 8-10, A-1040 Vienna, Austria}
\end{center}

\begin{center}
\begin{abstract}

We describe the quark gluon plasma (QGP) as a thermalized quark--gluon
system, the thermalized QGP phase of QCD. The hadronization of the
thermalized QGP phase is given in a way resembling a coalescence model
with correlated quarks and anti--quarks.  The input parameters of the
approach are the spatial volumes of the hadronization. We introduce
three dimensionless parameters $C_M$, $C_B$ and $C_{\bar{B}}$ related
to the spatial volumes of the production of low--lying mesons ($M$),
baryons ($B$) and antibaryons ($\bar{B}$). We show that at the
temperature $T= 175\, {\rm MeV}$ our predictions for the ratios of
multiplicities agree good with the presently available set of hadron
ratios measured for various experiments given by NA44, NA49, NA50 and
WA97 Collaborations on Pb+Pb collisions at 158\,GeV/nucleon, NA35
Collaboration on S+S collisions and NA38 Collaboration on O+U and S+U
collisions at 200\,GeV/nucleon.\\
\vspace{0.2in}
\end{abstract}
\end{center}

\newpage

\section{Introduction}
\setcounter{equation}{0}

\hspace{0.2in} Recently [1] we have suggested some kind of a
coalescence model [2] for the description of the hadronization from
the quark--gluon plasma (QGP) phase of QCD considered as a thermalized
quark--gluon system [3] at high densities and temperature in which
quarks, antiquarks and gluons being at the deconfined phase collide
frequently each other.  There is a belief [4] that the QGP phase of
the quark--gluon system can be realized in ultra--relativistic
heavy--ion collision ($E_{\rm cms}/{\rm nucleon} \gg 1\,{\rm GeV}$)
experiments.

At very high
energies of heavy--ion collisions the quark--gluon system is composed
from highly relativistic and very dense quarks, antiquarks and
gluons. By virtue of the asymptotic freedom the particles are almost
at liberty and due to high density collide themselves frequently that
leads to an equilibrium state. If to consider such a state as a
thermalized QGP phase of QCD, the probabilities of light
massless quarks $n_q(\vec{p}\,)$ and light massless antiquarks
$n_{\bar{q}}(\vec{p}\,)$, where $q = u$ or $d$, to have a momentum $p$
at a temperature $T$, can be described by the Fermi--Dirac
distribution functions [3,5]:
\begin{eqnarray}\label{label1.1}
n_q(\vec{p}\,) = \frac{1}{\textstyle e^{\textstyle - \nu(T) +p/T} +
1}\quad,\quad
n_{\bar{q}}(\vec{p}\,) = \frac{1}{\textstyle e^{\textstyle \nu(T) +
p/T} + 1},
\end{eqnarray}
where a temperature $T$ is measured in ${\rm MeV}$, $\nu(T) =
\mu(T)/T$, $\mu(T)$ is a chemical potential of the light massless
quarks $q = u,d$, depending on a temperature $T$ [5]. A chemical
potential of light antiquarks amounts to $- \mu(T)$. A positively
defined $\mu(T)$ provides an abundance of light quarks with respect to
light antiquarks for a thermalized state [1,4]. A chemical potential
$\mu(T)$ is a phenomenological parameter of the approach which we
would fix below [1]. 

The probability for gluons to have a momentum $\vec{p}$ at a
temperature $T$ is given by the Bose--Einstein distribution function
\begin{eqnarray}\label{label1.2}
n_g(\vec{p}\,) = \frac{1}{\textstyle e^{\textstyle p/T} - 1}.
\end{eqnarray}
Since a strangeness of the colliding heavy--ions amounts to zero,
the densities of strange quarks and antiquarks should be equal. The
former implies a zero--value of a chemical potential $\mu_s =
\mu_{\bar{s}} = 0$.  In this case the probabilities of strange quarks
and antiquarks can be given by
\begin{eqnarray}\label{label1.3}
n_s(\vec{p}\,) =
n_{\bar{s}}(\vec{p}\,) = \frac{1}{\textstyle e^{\textstyle
\sqrt{\vec{p}^{\,\,2} + m^2_s}/T} + 1},
\end{eqnarray}
where $m_s =135\,{\rm MeV}$ [6] is the mass of the strange quark and
antiquark. The value of the current $s$--quark mass $m_s =135\,{\rm
MeV}$ has been successfully applied to the calculation of chiral
corrections to  the amplitudes of low--energy interactions, form factors
and mass spectra of low--lying hadrons [7] and charmed heavy--light mesons
[8]. Unlike the massless antiquarks $\bar{u}$ and $\bar{d}$ for
which the suppression is caused by a chemical potential $\mu(T)$, the
strange quarks and antiquarks are suppressed by virtue of the
non--zero mass $m_s$.

In Ref.\,[1] we have supposed that a chemical potential $\mu(T)$, a
phenomenological parameter of the description of the QGP state as a
thermalized quark--gluon system at a temperature $T$, is an intrinsic
characteristic of a thermalized quark--gluon system. Thereby, if the
QGP is an excited state of the QCD vacuum, so a chemical potential
should exist not only for ultra--relativistic heavy--ion
collisions. Quark distribution functions of a thermalized quark--gluon
system at a temperature $T$ should be characterized by a chemical
potential $\mu(T)$ for any external state and any external conditions.
Since any state of a thermalized system is closely related to external
conditions, in order to obtain $\mu(T)$ we need only to specify the
external conditions of a thermalized quark--gluon system the
convenient for the determination of $\mu(T)$. 

Indeed, it is well--known [5] that the Helmholtz free energy
$F(T,V,N)$ defining the partition function $Z(T,V,N)$, $F(T,V,N) =
- \,T\,{\ell n}\,Z(T,V,N)$ which plays a central role in studying
thermalized systems, is nothing more than a work for an isothermic
process. Therefore, by producing external conditions keeping $T =
const$ and measuring a work one can get a full information about the
Helmholtz free energy $F_{\exp}(T,V,N)$. Then, in terms of this
Helmholtz free energy $F_{\exp}(T,V,N)$ one can obtain the partition
function $Z_{\exp}(T,V,N)$ which can be applied to the description of
the thermalized system at any $T$.

Following this idea we have fixed the chemical potential $\mu(T)$
in the form [1]:
\begin{eqnarray}\label{label1.4}
\frac{\mu(T)}{\mu_0} = \left[\frac{1}{2} + \frac{1}{2}\sqrt{1 +
\frac{4\pi^6}{27}\,\Bigg(\frac{T}{\mu_0}\Bigg)^6}\right]^{1/3} -
\left[-\frac{1}{2} + \frac{1}{2}\sqrt{1 +
\frac{4\pi^6}{27}\,\Bigg(\frac{T}{\mu_0}\Bigg)^6}\right]^{1/3}.
\end{eqnarray}
In the low--temperature limit $T \to 0$ we get
\begin{eqnarray}\label{label1.5}
\mu(T) = \mu_0\,\Bigg[1 - \frac{\pi^2}{3}\frac{T^2}{\mu^2_0} +
O\Big(T^6\Big)\Bigg].
\end{eqnarray}
where $\mu_0 = \mu(0) = 250\,{\rm MeV}$ is a chemical potential at
zero temperature [1].  The $T$--dependence of a chemical potential
given by Eq.(\ref{label1.5}) differs by a factor $1/4$ from the
low--temperature behaviour of a chemical potential of a thermalized
electron gas [9]. The former is caused by the contribution of
antiquarks.

In the high--temperature limit $T \to \infty$ a chemical potential
$\mu(T)$ defined by Eq.(\ref{label1.4}) drops like $T^{-2}$:
\begin{eqnarray}\label{label1.6}
\mu(T) = \frac{\mu^3_0}{\pi^2}\,\frac{1}{T^2} + O\Big(T^{-7}\Big).
\end{eqnarray}
A chemical potential drops very swiftly when a temperature increases.
Indeed, at $T = 160\,{\rm MeV}$ we obtain $\mu(T) \simeq \mu_0/4$,
while at $T = \mu_0$ a value of a chemical potential makes up about
tenth part of $\mu_0$, i.e. $\mu(T) \simeq \mu_0/10$. This implies
that at very high temperatures the function $\nu(T) = \mu(T)/T$
becomes small and the contribution of a chemical potential of light
quarks and antiquarks can be taken into account perturbatively. This
assumes in particular that at temperatures $T \ge \mu_0 = 250\,{\rm
MeV}$ the number of light antiquarks will not be suppressed by a
chemical potential relative to the number of light quarks.

In our approach the multiplicities of hadron production we define in
terms of quark and anti--quark distributions functions in a way
similar to a simple coalescence model [2] but for correlated quarks
and anti--quarks.  Indeed, in a coalescence model quarks and
anti--quarks are uncorrelated [2]. This allows to introduce separately
the number of light quarks $q$ and light anti--quarks $\bar{q}$ and
the number of strange quarks $s$ and strange anti--quarks $\bar{s}$
[2]. The subsequent calculation of multiplicities of hadrons in a
simple coalescence model resembles a quark counting. In fact,
multiplicities of hadrons are proportional to products of the number
of quarks $(q,s)$ and anti--quarks $(\bar{q}, \bar{s})$ in accord the
naive quark structure of hadrons. Since quarks and anti--quarks do not
correlate, so that the multiplicities of hadrons turn out to be
independent on the momenta of hadrons\,\footnote{The numbers of quarks
$(q,s)$ and anti--quarks $(\bar{q}, \bar{s})$ and the coefficients of
proportionality are free parameters of a simple coalescence model.
Therefore, a simple coalescence model contains seven free
parameters. Five of them can be fixed from experimental data [2].}.

In our approach the multiplicities of hadrons produced from the QGP
phase are described by momentum integrals of quark and anti--quark
distribution functions. Unlike a coalescence model these integrals
depend explicitly on the momenta of hadrons, temperature $T$ and
chemical potential $\mu(T)$.

For example, the multiplicities of the production of the $K^{\pm}$
mesons, $N_{K^{\pm}}(\vec{q}, T)$, and $\pi^{\pm}$ mesons,
$N_{\pi^{\pm}}(\vec{q}, T)$, are defined in our approach as follows
[1]:
\begin{eqnarray}\label{label1.7}
\hspace{-0.3in}N_{K^+}(\vec{q},T) &=&3\times 
\frac{C_M}{(M_K F_K)^{3/2}}\int\frac{d^3p}{(2\pi)^3}\,
\frac{1}{\textstyle
e^{\textstyle - \nu(T) +|\vec{p} - \vec{q}\,|/T} +
1}\frac{1}{\textstyle e^{\textstyle \sqrt{\vec{p}^{\,\,2} + m^2_s}/T}
+ 1},\nonumber\\
\hspace{-0.3in}N_{K^-}(\vec{q},T) &=&3\times 
\frac{C_M}{(M_K F_K)^{3/2}}\int\frac{d^3p}{(2\pi)^3}\,
\frac{1}{\textstyle e^{\textstyle \nu(T) + |\vec{p} - \vec{q}\,|/T}
+ 1}\frac{1}{\textstyle e^{\textstyle \sqrt{\vec{p}^{\,\,2} +
m^2_s}/T} + 1},\nonumber\\
\hspace{-0.3in}N_{\pi^{\pm}}(\vec{q},T) &=&3\times 
\frac{C_M}{(M_{\pi} F_{\pi})^{3/2}}
\int\frac{d^3p}{(2\pi)^3}\,\frac{1}{\textstyle e^{\textstyle -
\nu(T) +|\vec{p} - \vec{q}\,|/T} + 1}\frac{1}{\textstyle
e^{\textstyle  \nu(T) + p/T} + 1},
\end{eqnarray}
where $\vec{p}$ is a relative momentum of quarks and anti--quarks
coalesced in to a meson with a 3--momentum $\vec{q}$ at a temperature
$T$.  Since the main contribution to the integrals comes from the
relative momenta of order $p\sim T$, so that quark and anti--quarks
coalesce at relative momenta of order $p\sim T$. This agrees with the
order of transversal momenta of hadrons, $q_{\perp} \sim 2\div3\,T$,
coupled in the center of mass frame of heavy--ion collisions.  The
factor $3$ corresponds the number of quark color degrees of freedom,
$M_K = 500\,{\rm MeV}$, $F_K = 160\,{\rm MeV}$, $M_{\pi} = 140\,{\rm
MeV}$ and $F_{\pi} = 131\,{\rm MeV}$ are the masses and the leptonic
coupling constants of the $K$ and $\pi$ mesons, respectively [10]. The
dimensionless parameter $C_M$ is a free parameter of the approach. It
is the same for all low--lying mesons.

The multiplicities of the vector meson production, for example,
such as $K^{*\pm}$ and $\rho^{\pm}$ we define as 
\begin{eqnarray}\label{label1.8}
N_{K^{*+}}(\vec{q},T)&=&3\times\frac{C_M}{(M_{K^*}F_K)^{3/2}}
\int\frac{d^3p}{(2\pi)^3}\, \frac{1}{\textstyle
e^{\textstyle - \nu(T) +|\vec{p} - \vec{q}\,|/T} +
1}\frac{1}{\textstyle e^{\textstyle \sqrt{\vec{p}^{\,\,2} + m^2_s}/T}
+ 1},\nonumber\\ N_{K^{*-}}(\vec{q},T)
&=&3\times\frac{C_M}{(M_{K^*}
F_K)^{3/2}}\int\frac{d^3p}{(2\pi)^3}\, \frac{1}{\textstyle
e^{\textstyle \nu(T) + |\vec{p} - \vec{q}\,|/T} +
1}\frac{1}{\textstyle e^{\textstyle \sqrt{\vec{p}^{\,\,2} + m^2_s}/T}
+ 1},\nonumber\\ N_{\rho^{\pm}}(\vec{q},T)
&=&3\times \frac{C_M}{(M_{\rho} F_{\pi})^{3/2}}
\int\frac{d^3p}{(2\pi)^3}\,\frac{1}{\textstyle e^{\textstyle - \nu(T)
+|\vec{p} - \vec{q}\,|/T} + 1}\frac{1}{\textstyle e^{\textstyle \nu(T)
+ p/T} + 1},
\end{eqnarray}
where $M_{K^*} = 892\,{\rm MeV}$ and $M_{\rho} = 770\,{\rm MeV}$ are
the masses of the $K^*$ and $\rho$ mesons, respectively [10].

In the case of baryons and antibaryons we suggest to define the
multiplicities by using the diquark--quark picture of baryons and
antibaryons.  For example, the multiplicities of the proton ($p$) and
antiproton ($\bar{p}$) we write in the form
\begin{eqnarray}\label{label1.9}
N_{p}(\vec{q},T)&=&\frac{3!}{3!}\times \frac{C_B}{(M_{p}
F_{\pi})^{3/2}}\nonumber\\
&\times& \int\frac{d^3p}{(2\pi)^3}\, \frac{1}{\textstyle
\Big(e^{\textstyle - \nu(T) +p/T} +
1\Big)^2}\frac{1}{\textstyle e^{\textstyle - \nu(T) +|\vec{p} -
\vec{q}\,|/T} + 1},\nonumber\\
N_{\bar{p}}(\vec{q},T)&=&\frac{3!}{3!}\times \frac{C_{\bar{B}}}{(M_{p}
F_{\pi})^{3/2}}\nonumber\\
&\times& \int\frac{d^3p}{(2\pi)^3}\, \frac{1}{\textstyle
\Big(e^{\textstyle \nu(T) +p/T} +
1\Big)^2}\frac{1}{\textstyle e^{\textstyle \nu(T) +|\vec{p} -
\vec{q}\,|/T} + 1},
\end{eqnarray}
where a momentum $\vec{p}$ has a meaning of a relative momentum of
three--quark (three--anti--quark) system, $M_{p} = 940\,{\rm MeV}$ is
the mass of the proton and antiproton. As well as in the meson case
the main contribution to the momentum integrals comes from the momenta
of order $p \sim T$ providing a coalescence of three quarks (three
anti--quarks) into baryons (anti--baryons) at the momenta of order $p
\sim T$. That is again of order of transversal momenta of the produced
hadrons, $q_{\perp} \sim (2\div 3)\,T$, coupled in the center of mass
frame of heavy--ion collisions. The factor $3!$ in the numerator is
related to the quark colour degrees of freedom and defined by
$\varepsilon_{ijk}\,\varepsilon^{ijk} = 3!$, where $i,j$ and $k$ are
colour indices and run over $i = 1,2,3$ each. In turn, in the
denominator the factor $3!$ takes into account the identity of three
light quarks $(qqq)$ and three antiquarks
$(\bar{q}\bar{q}\bar{q}\,)$. In the isotopical limit we do not
distinguish $u$ and $d$ quarks as well as $\bar{u}$ and $\bar{d}$
antiquarks. The dimensionless parameters $C_B$ and $C_{\bar{B}}$ are
free parameters of the approach. Each of them is equal for all
components of octets of baryons and antibaryons, respectively, but
$C_B \not= C_{\bar{B}}$.

The paper is organized as follows. In Sect.\,2 we calculate the
theoretical values of multiplicities of the hadron production from
the thermalized QGP phase. The theoretical predictions and
experimental data are adduced in Table 1. In the Conclusion we discuss
the obtained results. A possible estimate of the absolute values of
our input parameters is discussed through the application of our
approach to the calculation of the number of baryons and antibaryons
relative to the number of photons at the early stage of the evolution
of the Universe assuming that this evolution goes through the
intermediate thermalized QGP phase.

\section{Multiplicities of hadron production from the thermalized QGP phase}
\setcounter{equation}{0}

\hspace{0.2in} Now let us proceed to the evaluation of multiplicities
of hadron production from the thermalized QGP phase of QCD. The
theoretical predictions for the different ratios of hadron
multiplicities we compare with experimental data adduced in Table I of
Ref.\,[11]. These are the data of various experiments given by NA44,
NA49, NA50 and WA97 Collaborations on Pb+Pb collisions at
158\,GeV/nucleon. Also we compare our results with the experimental
data obtained by NA35 Collaboration on S+S collisions and NA38
Collaboration on O+U and S+U collisions at 200\,GeV/nucleon. From
Table 1 of this paper one can see that in the whole the experimental
data for the hadron production are obtained for rapidities ranging
over the region $2.3 \le y \le 4.1$.  The relation between a
3--momentum $q$ and a rapidity $y$ reads
\begin{eqnarray}\label{label2.1}
q = \sqrt{M^2\,{\rm sh}^2y + q^2_{\perp}\,{\rm ch}^2y} \ge  M\,{\rm sh}y,
\end{eqnarray}
where $M$ and $\vec{q}_{\perp}$ are the mass and the transversal
momentum of the produced hadron.  For rapidities $y \in [2.3, 4.1]$
we get
\begin{eqnarray}\label{label2.2}
q  \ge  (5.0 \div 30.2)\,M.
\end{eqnarray}
Thus, for $K$ mesons and hadrons heavier than $K$ mesons typical
momenta are of order of 2.5\,GeV and greater. This gives a possibility
to investigate the momentum integrals defining multiplicities of the
hadron production at $q \to \infty$. As has been shown in Ref.\,[1]
the ratios of the multiplicities $R_{K^+K^-}(q, T) =
N_{K^+}(\vec{q},T)/N_{K^-}(\vec{q},T)$ and $R_{K^+\pi^+}(q, T) =
N_{K^+}(\vec{q},T)/N_{\pi^+}(\vec{q},T)$ are smooth functions of $q$,
wobbling slightly around the asymptotic values obtained at $q \to
\infty$, and describe good the experimental data at $T = 175\,{\rm
MeV}$. Below the theoretical results on the multiplicities of the
hadron production we would compare with experimental data at $T =
175\,{\rm MeV}$. In this case it is obvious that the typical momenta
of the momentum integrals defining the multiplicities of hadron
production are of order $p\sim T$. Therefore, in our approach the
momenta at which quarks and anti--quarks coalesces into hadrons should
be of order $p\sim T$. This agrees with the order of transversal
momenta of hadrons, $q_{\perp} \sim 2\div3\,T$, coupled in the center
of mass frame of heavy--ion collisions.

Hence, for rapidities ranging over the region $2.3 \le y \le 4.1$ the
typical total 3--momenta of hadrons are greater than 1\,GeV. Thereby,
multiplicities of the hadron production can be calculated in the
asymptotic regime at $q \gg T$.

We would like to accentuate that the relative momenta at which quarks
and anti--quarks coalesce into hadrons are described effectively by
the momentum of integration $\vec{p}$.

At $q \gg T$ the multiplicities of the hadron production defined by
Eqs.(\ref{label1.7})--(\ref{label1.9}) can be represented in the
following form
\begin{eqnarray}\label{label2.3}
N_{\pi^+}(\vec{q},T)
&=&N_{\pi^-}(\vec{q},T)= N_{\pi^0}(\vec{q},T) = \frac{3\,C_M}{(M_{\pi}
F_{\pi})^{3/2}}\,e^{\textstyle - q/T}\,I_{\pi}(T),\nonumber\\
N_{K^+}(\vec{q},T) &=&N_{K^0}(\vec{q},T)=\frac{3\,C_M}{(M_K
F_K)^{3/2}}\,e^{\textstyle - q/T}\,e^{\textstyle +
{\nu}(T)}\,I_{K}(T),\nonumber\\ N_{K^-}(\vec{q},T)
&=&N_{\bar{K}^0}(\vec{q},T)=\frac{3\,C_M}{(M_K
F_K)^{3/2}}\,e^{\textstyle - q/T}\,e^{\textstyle {-
\nu}(T)}\,I_{\bar{K}}(T),\nonumber\\
N_{K^0_S}(\vec{q},T)&=& \frac{1}{2}\,N_{K^0}(\vec{q},T)+
\frac{1}{2}\,N_{\bar{K}^0}(\vec{q},T) =\frac{1}{2}\,\frac{3\,C_M}{(M_K
F_K)^{3/2}}\,e^{\textstyle - q/T}\,e^{\textstyle
+\nu(T)}\,I_{K^0_S}(T),\nonumber\\
N_{\eta}(\vec{q},T)&=& \sin^2\bar{\theta}\,
\frac{3\,C_M}{(M_{\eta} F_{\pi})^{3/2}}\,e^{\textstyle
-\,q/T}\,I_{\pi}(T) + \cos^2\bar{\theta}\,\frac{3\, C_M}{(M_{\eta} F_S)^{3/2}\,}e^{\textstyle
-\,q/T}\,I_{\eta}(T),\nonumber\\
N_{\phi}(\vec{q},T)&=& \frac{3\, C_M}{(M_{\phi}F_S)^{3/2}}
\,e^{\textstyle -\,q/T}\,I_{\phi}(T),\nonumber\\
N_p(\vec{q},T)&=& \frac{C_B}{(M_{p} F_{\pi})^{3/2}}\,e^{\textstyle -
q/T}\,e^{\textstyle +\,\nu(T)}\,I_p(T),\nonumber\\
N_{\Lambda}(\vec{q},T)&=&
\frac{3\, C_B}{(M_{\Lambda}
F_K)^{3/2}}\,e^{\textstyle - q/T}\,e^{\textstyle
+2\,\nu(T)}\,I_{\Lambda}(T),\nonumber\\
N_{\Xi}(\vec{q},T)&=& \frac{3\, C_B}{(M_{\Xi}
F_S)^{3/2}}\,e^{\textstyle - q/T}\,e^{\textstyle
+\nu(T)}\,I_{\Xi}(T),\nonumber\\
N_{\Omega}(\vec{q},T)&=&\frac{C_B}{(M_{\Omega}
F_S)^{3/2}}\,e^{\textstyle - q/T}\,I_{\Omega}(T),\nonumber\\
N_{\bar{p}}(\vec{q},T)&=& \frac{C_{\bar{B}}}{(M_{p}
F_{\pi})^{3/2}}\,e^{\textstyle - q/T}\,e^{\textstyle -
\nu(T)}\,I_{\bar{p}}(T),\nonumber\\
N_{\bar{\Lambda}}(\vec{q},T) &=&
\frac{3\, C_{\bar{B}}}{(M_{\Lambda}
F_K)^{3/2}}\,e^{\textstyle - q/T}\,e^{\textstyle
-\,2\,\nu(T)}\,I_{\bar{\Lambda}}(T),\nonumber\\
N_{\bar{\Xi}}(\vec{q},T)&=& \frac{3\, C_{\bar{B}}}{(M_{\Xi}
F_S)^{3/2}}\,e^{\textstyle - q/T}\,e^{\textstyle
- \nu(T)}\,I_{\bar{\Xi}}(T),\nonumber\\
N_{\bar{\Omega}}(\vec{q},T) &=& 
\frac{C_{\bar{B}}}{(M_{\Omega}
F_S)^{3/2}}\,e^{\textstyle - q/T}\,I_{\bar{\Omega}}(T),
\end{eqnarray}
where the structure functions $I_i(T)\,(i= \pi, K, \bar{K}, \ldots)$ are
defined by
\begin{eqnarray}\label{label2.4}
\hspace{-0.5in}&&I_{\pi}(T) = e^{\textstyle +
\nu(T)}\int\limits^{\infty}_0\frac{dp}{4\pi^2}\, \frac{p^2}{\textstyle
e^{\textstyle + \nu(T) + p/T} + 1} + \,e^{\textstyle -
\nu(T)}\int\limits^{\infty}_0\frac{dp}{4\pi^2}\, \frac{p^2}{\textstyle
e^{\textstyle - \nu(T) + p/T} + 1}=\nonumber\\
\hspace{-0.5in}&&=\frac{T^3}{4\pi^2}\,\Bigg[e^{\textstyle +
\nu(T)}\int\limits^{\infty}_0\frac{dx\,x^2}{\textstyle e^{\textstyle +
\nu(T) + x} + 1} + \,e^{\textstyle -
\nu(T)}\int\limits^{\infty}_0\frac{dx\,x^2}{\textstyle e^{\textstyle -
\nu(T) + x} + 1}\Bigg] =  3.153\,\frac{T^3}{4\pi^2},\nonumber\\
\hspace{-0.5in}&&I_{K}(T)=\int\limits^{\infty}_0\frac{dp}{4\pi^2}\,
\frac{p^2}{\textstyle e^{\textstyle \sqrt{p^{\,2} + m^2_s}/T} + 1} +\,
e^{\textstyle - \nu(T)}\int\limits^{\infty}_0\frac{dp}{4\pi^2}\,
\frac{p^2}{\textstyle e^{\textstyle - \nu(T) + p/T} + 1} =\nonumber\\
\hspace{-0.5in}&&= \frac{T^3}{4\pi^2}\,
\Bigg[\frac{m^3_s}{T^3}\int\limits^{\infty}_0\frac{dx\,x^2}{\textstyle
e^{\,\textstyle (m_s/T)\sqrt{1 + x^2}} + 1} + e^{\textstyle -
\nu(T)}\int\limits^{\infty}_0\frac{dx\,x^2}{\textstyle e^{\textstyle -
\nu(T) + x} + 1}\Bigg] = 2.924\,\frac{T^3}{4\pi^2},\nonumber\\
\hspace{-0.5in}&&I_{\bar{K}}(T) = \int\limits^{\infty}_0\frac{dp}{4\pi^2}\,
\frac{p^2}{\textstyle e^{\textstyle \sqrt{p^{\,2} + m^2_s}/T} + 1} +\,
e^{\textstyle + \nu(T)}\int\limits^{\infty}_0\frac{dp}{4\pi^2}\,
\frac{p^2}{\textstyle e^{\textstyle + \nu(T) + p/T} + 1} =\nonumber\\
\hspace{-0.5in}&&=\frac{T^3}{4\pi^2}\Bigg[\frac{m^3_s}{T^3}\,\int\limits^{\infty}_0\frac{dx\,x^2}{\textstyle
e^{\textstyle (m_s/T)\sqrt{1 + x^2}} + 1} + e^{\textstyle +
\nu(T)}\int\limits^{\infty}_0\frac{dx\,x^2}{\textstyle e^{\textstyle +
\nu(T) + x} + 1}\Bigg] = 3.463\,\frac{T^3}{4\pi^2},\nonumber\\
\hspace{-0.5in}&&I_{\eta}(T) = I_{\phi}(T)= \nonumber\\
\hspace{-0.5in}&&=\int\limits^{\infty}_0\frac{dp}{2\pi^2}\,
\frac{p^2}{\textstyle e^{\textstyle \sqrt{\vec{p}^{\,\,2} + m^2_s}/T}
+ 1} =
\frac{m^3_s}{2\pi^2}\int\limits^{\infty}_0\frac{d\,x^2}{\textstyle
e^{\textstyle (m_s/T)\sqrt{1 + x^2}} + 1} =
3.522\,\frac{m^3_s}{2\pi^2},\nonumber\\
\hspace{-0.5in}&&I_p(T)= \int\limits^{\infty}_0\frac{dp}{2\pi^2}\, \frac{1}{\textstyle
\Big(e^{\textstyle - \nu(T) +p/T} + 1\Big)^2} =
\frac{T^3}{2\pi^2}\int\limits^{\infty}_0\frac{dx\,x^2}{\textstyle
\Big(e^{\textstyle - \nu(T) + x} + 1\Big)^2} =
0.253\,\frac{T^3}{2\pi^2},\nonumber\\
\hspace{-0.5in}&&I_{\Lambda}(T) =  e^{\textstyle - \nu(T)}\int\limits^{\infty}_0\frac{dp}{4\pi^2}\,\frac{p^2}{\textstyle e^{\textstyle\sqrt{p^{\,2} + m^2_s}/T} + 1}\,\frac{1}{\textstyle
e^{\textstyle - \nu(T) +p/T} + 1}\nonumber\\
\hspace{-0.5in}&&+\, e^{\textstyle -
2\,\nu(T)}\int\limits^{\infty}_0\frac{dp}{4\pi^2}\,\frac{p^2}{\textstyle
\Big(e^{\textstyle - \nu(T) +p/T} + 1\Big)^2}=\nonumber\\
\hspace{-0.5in}&&=\frac{m^3_s}{4\pi^2}\Bigg[e^{\textstyle - \nu(T)}\int\limits^{\infty}_0\frac{dx\,x^2}{\textstyle e^{\textstyle(m_s/T)\sqrt{1 + x^2}} + 1}
\,\frac{1}{\textstyle e^{\textstyle - \nu(T)
+(m_s/T)\,x} + 1}\nonumber\\
\hspace{-0.5in}&&+ e^{\textstyle -\,2\, \nu(T)}\int\limits^{\infty}_0\frac{dx\,x^2}{\textstyle
\Big(e^{\textstyle - \nu(T) +(m_s/T)\,x} + 1\Big)^2}\Bigg]= 0.582\,\frac{m^3_s}{4\pi^2},\nonumber\\
\hspace{-0.5in}&&I_{\Xi}(T)=\int\limits^{\infty}_0\frac{dp}{4\pi^2}\,
\frac{p^2}{\textstyle 
\Big(e^{\textstyle\sqrt{p^{\,2} + m^2_s}/T} + 1\Big)^2}\nonumber\\
\hspace{-0.5in}&&+\,
e^{\textstyle - \nu(T)}\int\limits^{\infty}_0
\frac{dp}{4\pi^2}\,\frac{p^2}{\textstyle 
e^{\textstyle\sqrt{p^{\,2} + m^2_s}/T} + 1}\,\frac{1}{\textstyle
e^{\textstyle - \nu(T) +p/T} + 1}=\nonumber\\
\hspace{-0.5in}&&=\frac{m^3_s}{4\pi^2}
\Bigg[\int\limits^{\infty}_0\frac{dx\,x^2}{\textstyle 
\Big(e^{\textstyle (m_s/T)\sqrt{1 + x^2}} + 1\Big)^2} + 
e^{\textstyle - \nu(T)}\int\limits^{\infty}_0
\frac{dx\,x^2}{\textstyle e^{\textstyle(m_s/T)\sqrt{1 + x^2}} + 1}
\nonumber\\
\hspace{-0.5in}&&\times \frac{1}{\textstyle e^{\textstyle - \nu(T)
+(m_s/T)\,x} + 1}\Bigg]= 0.528\,\frac{m^3_s}{4\pi^2}, \nonumber\\
\hspace{-0.5in}&&I_{\Omega}(T)= \int\limits^{\infty}_0\frac{dp}{2\pi^2}\,
\frac{p^2}{\textstyle 
\Big(e^{\textstyle\sqrt{p^{\,2} + m^2_s}/T} + 1\Big)^2}=\frac{m^3_s}{2\pi^2}\int\limits^{\infty}_0\frac{dx\,x^2}{\textstyle 
\Big(e^{\textstyle (m_s/T)\sqrt{1 + x^2}} + 1\Big)^2}=\nonumber\\
&&=0.251\,\frac{m^3_s}{2\pi^2},\nonumber\\
\hspace{-0.5in}&&I_{\bar{p}}(T) = \int\limits^{\infty}_0\frac{dp}{2\pi^2}\,
 \frac{p^2}{\textstyle \Big(e^{\textstyle \nu(T) +p/T} + 1\Big)^2} =
 \frac{T^3}{2\pi^2}\int\limits^{\infty}_0 \frac{dx\,x^2}{\textstyle
 \Big(e^{\textstyle \nu(T) + x} + 1\Big)^2} =
 0.097\,\frac{T^3}{2\pi^2},\nonumber\\
\hspace{-0.5in}&&\times \frac{1}{\textstyle e^{\textstyle + \nu(T)
+(m_s/T)\,x} + 1}\Bigg]= 0.558\,\frac{m^3_s}{4\pi^2}, \nonumber\\
\hspace{-0.5in}&&I_{\bar{\Lambda}}(T)= e^{\textstyle +
\nu(T)}\int\limits^{\infty}_0\frac{dp}{4\pi^2}\,\frac{p^2}{\textstyle
e^{\textstyle\sqrt{p^{\,2} + m^2_s}/T} + 1}\,\frac{1}{\textstyle
e^{\textstyle + \nu(T) +p/T} + 1}\nonumber\\
\hspace{-0.5in}&&+\, e^{\textstyle +
2\,\nu(T)}\int\limits^{\infty}_0\frac{dp}{4\pi^2}\,\frac{p^2}{\textstyle
\Big(e^{\textstyle + \nu(T) +p/T} + 1\Big)^2}=\nonumber\\
\hspace{-0.5in}&=&\frac{m^3_s}{4\pi^2}\Bigg[e^{\textstyle +
\nu(T)}\int\limits^{\infty}_0\frac{dx\,x^2}{\textstyle
e^{\textstyle(m_s/T)\sqrt{1 + x^2}} + 1} \,\frac{1}{\textstyle
e^{\textstyle + \nu(T) +(m_s/T)\,x} + 1}\nonumber\\
\hspace{-0.5in}&&+\, e^{\textstyle +\,2\, 
\nu(T)}\int\limits^{\infty}_0\frac{dx\,x^2}{\textstyle
\Big(e^{\textstyle + 
\nu(T) +(m_s/T)\,x} + 1\Big)^2}\Bigg]= 
0.686\,\frac{m^3_s}{4\pi^2},\nonumber\\
\hspace{-0.5in}&&I_{\bar{\Xi}}(T)= \int\limits^{\infty}_0\frac{dp}{4\pi^2}\,
\frac{p^2}{\textstyle 
\Big(e^{\textstyle\sqrt{p^{\,2} + m^2_s}/T} + 1\Big)^2}\nonumber\\
\hspace{-0.5in}&&+\,
e^{\textstyle +\nu(T)}\int\limits^{\infty}_0
\frac{dp}{4\pi^2}\,\frac{p^2}{\textstyle 
e^{\textstyle\sqrt{p^{\,2} + m^2_s}/T} + 1}\,\frac{1}{\textstyle
e^{\textstyle + \nu(T) +p/T} + 1}=\nonumber\\
\hspace{-0.5in}&&=\frac{m^3_s}{4\pi^2}
\Bigg[\int\limits^{\infty}_0\frac{dx\,x^2}{\textstyle 
\Big(e^{\textstyle (m_s/T)\sqrt{1 + x^2}} + 1\Big)^2} + \,
e^{\textstyle + \nu(T)}\int\limits^{\infty}_0
\frac{dx\,x^2}{\textstyle e^{\textstyle(m_s/T)\sqrt{1 + x^2}} + 1}\nonumber\\
\hspace{-0.5in}&&\times \frac{1}{\textstyle e^{\textstyle + \nu(T)
+(m_s/T)\,x} + 1}\Bigg]= 0.558\,\frac{m^3_s}{4\pi^2}, \nonumber\\
\nonumber\\
\hspace{-0.5in}&&I_{\bar{\Omega}}(T) = I_{\Omega}(T).
\end{eqnarray}
The numerical values of the integrals are obtained at $m_s =
135\,{\rm MeV}$ and $T = 175\,{\rm MeV}$.

The theoretical ratios of multiplicities of the hadron production
which we compare with measured experimentally we define as follows
\begin{eqnarray}\label{label2.5}
R_{K^+K^-}(q,T) &=&\frac{N_{K^+}(\vec{q},T)}{N_{K^-}(\vec{q},T)} =
e^{\textstyle +\,2\,\nu(T)}\,\frac{I_{K}(T)}{I_{\bar{K}}(T)} =
1.520,\nonumber\\ R_{K^+\pi^+}(q,T)&=&
\frac{N_{K^+}(\vec{q},T)}{N_{\pi^+}(\vec{q},T)} =
\Bigg(\frac{M_{\pi}F_{\pi}}{M_KF_K}\Bigg)^{3/2}\, e^{\textstyle
+\,{\nu}(T)}\,\frac{I_{K}(T)}{I_{\pi}(T)} = 0.139,\nonumber\\
R_{K^-\pi^-}(q,T)&=&
\frac{N_{K^-}(\vec{q},T)}{N_{\pi^-}(\vec{q},T)} =
\Bigg(\frac{M_{\pi}F_{\pi}}{M_KF_K}\Bigg)^{3/2}\, e^{\textstyle
-\,{\nu}(T)}\,\frac{I_{\bar{K}}(T)}{I_{\pi}(T)} = 0.090,\nonumber\\
R_{K^0_S\pi^-}(q,T)&=&\frac{N_{K^0_S}(\vec{q},T)}{N_{\pi^-}(\vec{q},T)}
= \frac{1}{2}\,\Bigg(\frac{M_{\pi}F_{\pi}}{M_KF_K}\Bigg)^{3/2}\,
e^{\textstyle +\,\nu(T)}\,\frac{I_{K^0_S}(T)}{I_{\pi^-}(T)}
=0.113,\nonumber\\ R_{\Xi\Lambda}(q,T) &=&
\frac{N_{\Xi}(\vec{q},T)}{N_{\Lambda}(\vec{q},T)} =
\Bigg(\frac{M_{\Lambda}F_K}{M_{\Xi}F_S}\Bigg)^{3/2}\, e^{\textstyle
-\, \nu(T)}\,\frac{I_{\Xi}(T)}{I_{\Lambda}(T)} = 0.108,\nonumber\\
R_{\Omega\Xi}(q,T)&=&\frac{N_{\Omega}(\vec{q},T)}{N_{\Xi}(\vec{q},T)}
= \frac{1}{3}\,\Bigg(\frac{M_{\Xi}}{M_{\Omega}}\Bigg)^{3/2}\,
e^{\textstyle - \,\nu(T)}\,\frac{I_{\Omega}(T)}{I_{\Xi}(T)} =
0.166,\nonumber\\ R_{\bar{\Lambda}\bar{p}}(q,T) &=&
\frac{N_{\bar{\Lambda}}(\vec{q},T)}{N_{\bar{p}}(\vec{q},T)} =
3\,\Bigg(\frac{M_{p}F_{\pi}}{M_{\Lambda}F_K}\Bigg)^{3/2}\,
e^{\textstyle - \,\nu(T)}\,\frac{I_{\bar{\Lambda}}(T)}{I_{\bar{p}}(T)}
= 2.081,\nonumber\\ R_{\bar{\Xi}\bar{\Lambda}}(q,T) &=&
\frac{N_{\bar{\Xi}}(\vec{q},T)}{N_{\bar{\Lambda}}(\vec{q},T)} =
\Bigg(\frac{M_{\Lambda}F_K}{M_{\Xi}F_S}\Bigg)^{3/2}\, e^{\textstyle
+\,\nu(T)}\,\frac{I_{\bar{\Xi}}(T)}{I_{\bar{\Lambda}}(T)}=
0.173,\nonumber\\ R_{\bar{\Omega}\bar{\Xi}}(q,T) &=&
\frac{N_{\bar{\Omega}}(\vec{q},T)}{N_{\bar{\Xi}}(\vec{q},T)} =
\frac{1}{3}\,\Bigg(\frac{M_{\Xi}}{M_{\Omega}}\Bigg)^{3/2}\,e^{\textstyle
+ \,\nu(T)}\,\frac{I_{\bar{\Omega}}(T)}{I_{\bar{\Xi}}(T)}
=0.282,\nonumber\\ R_{\bar{\Omega}\Omega}(q,T) &=&
\frac{N_{\bar{\Omega}}(\vec{q},T)}{N_{\Omega}(\vec{q},T)} =
\frac{C_{\bar{B}}}{C_B} =R^{\exp}_{\bar{\Omega}\Omega} = 0.46 \pm
0.15,\nonumber\\ R_{\bar{\Lambda}\Lambda}(q,T) &=&
\frac{N_{\bar{\Lambda}}(\vec{q},T)}{N_{\Lambda}(\vec{q},T)} =
\frac{C_{\bar{B}}}{C_B}\times e^{\textstyle
-\,4\,\nu(T)}\,\frac{I_{\bar{\Lambda}}(T)}{I_{\Lambda}(T)} = 0.168\pm
0.055,\nonumber\\ R_{\bar{\Xi}\Xi}(q,T) &=&
\frac{N_{\bar{\Xi}}(\vec{q},T)}{N_{\Xi}(\vec{q},T)} =
\frac{C_{\bar{B}}}{C_B}\times e^{\textstyle
-\,2\,\nu(T)}\,\frac{I_{\bar{\Xi}}(T)}{I_{\Xi}(T)} = 0.270\pm
0.088,\nonumber\\ R_{\eta \pi^0}(q,T)&=&
\frac{N_{\eta}(\vec{q},T)}{N_{\pi^0}(\vec{q},T)}
=\sin^2\bar{\theta}\,\Bigg(\frac{M_{\pi}}{M_{\eta}}\Bigg)^{3/2} +
\cos^2\bar{\theta}\,
\Bigg(\frac{M_{\pi}F_{\pi}}{M_{\eta}F_S}\Bigg)^{3/2}\,\frac{I_{\eta}(T)}{I_{\pi^+}(T)}=
0.088,\nonumber\\ R_{\phi\pi}(q,T)&=&
\frac{N_{\phi}(\vec{q},T)}{N_{\pi}(\vec{q},T)} =
\Bigg(\frac{M_{\pi}F_{\pi}}{M_{\phi}F_S}\Bigg)^{3/2}\,
\frac{I_{\phi}(T)}{I_{\pi}(T)} = 7.84\times 10^{-3},\nonumber\\
R_{\phi(\rho^0 + \omega^0)}(q,T)&=&
\frac{N_{\phi}(\vec{q},T)}{N_{\rho^0}(\vec{q},T) + N_{\omega^0}(\vec{q},T)} =
\Bigg(\frac{M_{\rho}F_{\pi}}{M_{\phi}F_S}\Bigg)^{3/2}\,
\frac{I_{\phi}(T)}{I_{\pi}(T)} = 0.103,\nonumber\\
R_{\phi K^0_S}(q,T) &=&
\frac{N_{\phi}(\vec{q},T)}{N_{K^0_S}(\vec{q},T)} = 2\,\Bigg(\frac{M_K
F_K}{M_{\phi}F_S}\Bigg)^{3/2}\,e^{\textstyle -\,\nu(T)}\,
\frac{I_{\phi}(T)}{I_{K^0_S}(T)} = 0.071,\nonumber\\ R_{\Lambda
K^0_S}(q,T) &=& \frac{N_{\Lambda}(\vec{q},T)}{N_{K^0_S}(\vec{q},T)} =
\frac{C_B}{C_M}\times 2\,e^{\textstyle
+\,\nu(T)}\,\frac{I_{\Lambda}(T)}{I_{K^0_S}(T)} =
\frac{C_B}{C_M}\times 0.148 =\nonumber\\ &=& R^{\exp}_{\Lambda K^0_S} =
0.65\pm 0.11 \to \frac{C_B}{C_M} = 4.39\pm 0.74, \nonumber\\
R_{p K^+}(q,T) &=& \frac{N_p(\vec{q},T)}{N_{K^+}(\vec{q},T)} =
\frac{C_B}{C_M}\times  \frac{1}{3}\,\Bigg(\frac{M_K F_K}{M_p
F_{\pi}}\Bigg)^{3/2}\frac{I_p(T)}{I_{K}(T)} =  0.136\pm 0.023,\nonumber\\
R_{\bar{p}K^-}(q,T) &=&
\frac{N_{\bar{p}}(\vec{q},T)}{N_{K^-}(\vec{q},T)} =
\frac{C_{\bar{B}}}{C_M}\times \frac{1}{3}\,\Bigg(\frac{M_K F_K}{M_p
F_{\pi}}\Bigg)^{3/2}\frac{I_{\bar{p}}(T)}{I_{\bar{K}}(T)} = 0.020 \pm 0.007.
\end{eqnarray}
The constant $F_S = 3.5\,F_{\pi}$ is related to the leptonic constant
of the pseudoscalar meson containing only $s$--quarks, $s\bar{s}$
[12]. We have estimated $F_S$ in agreement with the experimental data
on the $\eta(550)/\pi^0$ and $\phi(1020)/\pi$ production. For the
description of the multiplicity of the $\eta(550)$ meson production we
have taken into account that the low--energy meson phenomenology [10,12,13]
gives the following quark structure of the $\eta(550)$ meson:
\begin{eqnarray}\label{label2.6}
\eta(550) = (q\bar{q})\,\sin\bar{\theta} +
(s\bar{s})\,\cos\bar{\theta},
\end{eqnarray}
where $\bar{\theta} = \vartheta_0 - \vartheta_P$ with
$\vartheta_0 = 35.264^0$, the ideal mixing angle, and $\vartheta_P$, the
octet--singlet mixing angle. Recent analysis of the value of the
octet--singlet mixing angle carried out by Bramon, Escribano and
Scadron [13] gives $\vartheta_P = -\,16.9\pm 1.7\,^0$. For the
$\phi(1020)$ meson we have supposed the $s\bar{s}$ quark structure [10,12].

\section{Conclusion}
\setcounter{equation}{0}

\hspace{0.2in} The theoretical and experimental values of the ratios
of hadron production are adduced in Table 1.  From Table 1 one can see
a good agreement between presently available set of hadron ratios
measured for various experiments given by NA44, NA49, NA50 and WA97
Collaborations on Pb+Pb collisions at 158\,GeV/nucleon, NA35
Collaboration on S+S collisions and NA38 Collaboration on O+U and S+U
collisions at 200\,GeV/nucleon and theoretical predictions for the
ratios of multiplicities of hadron production from the thermalized QGP
phase at a temperature $T = 175\,{\rm MeV}$. Save the ratio
$\bar{\Lambda}/\bar{p}$, $(\bar{\Lambda}/\bar{p})_{\rm th} = 2.081$
and $(\bar{\Lambda}/\bar{p})_{\exp} = 3\pm 1$, the theoretical results
agree with the experimental ones with accuracy better than
18$\%$. 

In the our approach multiplicities of hadron production are defined by
momentum integrals on quark (antiquark) distribution functions in
accordance with the phenomenological quark structure of the hadron.
For the analysis of the multiplicities of the baryon and antibaryon
production in terms of the quark and antiquark distribution functions
we have followed the diquark--quark picture for baryons and
antibaryons. This has allowed to describe multiplicities of the
baryon, antibaryon and meson production on the same footing.

For the explanation of experimental data on the hadron production in
ultrarelativistic heavy--ion collisions we have used three input
parameters $C_{\bar{B}}/C_B$, $C_M/C_B$ and $F_S$. These parameters
are related to the spatial volumes of hadronization of the quarks and
antiquarks from the thermalized QGP phase. The first two parameters
have been fixed from the experimental data on the ratios
$(\bar{\Omega}/\Omega)_{\exp} = 0.46\pm 0.15$, $(\Lambda/K^0_S)_{\exp}
= 0.65\pm 0.11$. This gives $C_{\bar{B}}/C_B = 0.46\pm 0.15$ and
$C_M/C_B = 0.23\pm 0.04$. In turn, the value of the parameter $F_S =
3.5\,F_{\pi} = 458.5\,{\rm MeV}$ is a result of a smooth fit of the
ratios of hadrons containing the $ss$ and $s\bar{s}$ components in the
quark structure. In the bulk our approach to the hadronization from
the QGP phase of QCD has succeeded in describing 21 experimental data
on ultrarelativistic heavy--ion collisions.

Unlike other approaches [2--4] the ratio of the $\bar{\Omega}$ and
$\Omega$ baryon production is an input parameter in our model
$C_{\bar{B}}/C_B$. The ratio $\bar{\Omega}/\Omega$ does not depend in
our approach on both the momenta of baryons and the temperature. The
former is due to the zero--value of the $s$--quark chemical potential,
$\mu_s = \mu_{\bar{s}} = 0$. As a result the ratio
$\bar{\Omega}/\Omega$ can be only fitted in our approach. By fitting
the ratio $\bar{\Omega}/\Omega$ from experimental data and applying
this value to the description of other ratios of the baryon and
antibaryon production for the thermalized QGP phase we have found a
good agreement with experimental data. This confirms a
self--consistency of the approach.

In turn, the distinction between the parameters $C_B$ and
$C_{\bar{B}}$ can be related to a well--known factor of the
baryon--antibaryon asymmetry in the Universe which one could put
phenomenologically at the early stage of the evolution of the Universe
whether the baryon synthesis in it goes through the intermediate QGP
phase. Indeed, as has been stated by B\"orner [29]: {\it Within the
standard big--bang model, however, there seems to be little chance of
achieving a physical separation of baryon and antibaryon phases in an
initially baryon--symmetric cosmological model. If the baryon number
was exactly conserved -- as it assumed to be the standard model -- the
small asymmetry necessary for our existence must be postulated
initially. Grand unified theories offer the possibility of creating
this small asymmetry from physical processes.}

In our approach the baryon--antibaryon asymmetry at the hadronic level
can be realized phenomenologically in terms of a different rate of
hadronization of baryons and antibaryons caused by the input parameter
$C_{\bar{B}}/C_B = 0.46\pm 0.15$ fixed through the experimental data
on the ratio $\bar{\Omega}/\Omega$ production in ultrarelativistic
heavy--ion collisions. For the total number of antibaryons
$N_{\bar{B}}$ relative to the total number of baryons $N_B$ produced
for the baryon--antibaryon synthesis at the early stage of the
evolution of the Universe at a temperature $T = 175\,{\rm MeV}$ [30] and
gone through the intermediate thermalized QGP phase we predict
\begin{eqnarray}\label{label3.1}
\frac{N_{\bar{B}}}{N_B} = 0.41\times \frac{C_{\bar{B}}}{C_B} = 0.19\pm
0.06.
\end{eqnarray}
This result can be supported by a trivial estimate in the
approximation of the equilibrium baryon and anti--baryon gases. In
such an approximation the ratio $N_{\bar{B}}/N_B$ is defined by
\begin{eqnarray}\label{label3.2}
\frac{N_{\bar{B}}}{N_B} = e^{\textstyle -2\mu_B(T)/T} = e^{\textstyle
-6\mu(T)/T} = 0.17,
\end{eqnarray}
where a chemical potential $\mu(T)$ is given by Eq.(\ref{label1.4})
and $T=175\,{\rm MeV}$.

Thus, at the early stage of the evolution of the Universe the number
of antibaryons should be of order of magnitude less compared with the
number of baryons, $N_{\bar{B}} \sim 0.2\,N_B$.  According to B\"orner
[31] it is more than enough for the existence of the life in the
Universe. Recall that the standard approach [31] predicts for every
$10^9$ antibaryons only $(10^9 + 1)$ baryons. As has been stated by
B\"orner: {\it It is to that one part in $10^9$ excess of ordinary
matter that we owe our existence}! [31].

For the early Universe the total number of baryons and antibaryons was
roughly equal to the number of photons  $N_{\rm ph}$ [31]:
\begin{eqnarray}\label{label3.3}
\frac{\displaystyle N_{\bar{B}} + N_B}{N_{\rm ph}}\simeq 1.
\end{eqnarray}
Since the density of photons is equal to [32]
\begin{eqnarray}\label{label3.4}
\frac{N_{\rm ph}}{V} =  \frac{2.404}{\pi^2}\,T^3,
\end{eqnarray}
where $V$ is the volume of the early Universe, and the density of the
total number of baryons and antibaryons $N_{\bar{B}} + N_B$ calculated
in our approach at $T = 175\,{\rm MeV}$ amounts to
\begin{eqnarray}\label{label3.5}
\frac{\displaystyle N_{\bar{B}} + N_B}{V} =  C_B\times 0.442\times T^3,
\end{eqnarray}
we can estimate the numerical value of the parameter $C_B$:
\begin{eqnarray}\label{label3.6}
C_B \simeq 0.55 \pm 0.08.
\end{eqnarray}
This gives the estimate of other input parameters $C_{\bar{B}}$ and
$C_M$:
\begin{eqnarray}\label{label3.7}
C_{\bar{B}} &\simeq& 0.25 \pm 0.08,\nonumber\\
C_M &\simeq& 0.13 \pm 0.03.
\end{eqnarray}
The analysis of the influence of the input parameter $C_{\bar{B}}/C_B
= 0.46\pm 0.15$ on the evolution of the baryon--antibaryon asymmetry
in the Universe from the early Universe up to the present epoch and
the formation of a dark and strange matter in the Universe [10] we are
planning to carry out in our forthcoming publications.

Now we would like to discuss in more details our approach when
compared with a simple coalescence model [2]. The main distinction of
our approach from a simple coalescence model is in the correlation
between quarks and anti--quarks coalescing into hadrons. In fact, in a
simple coalescence model quarks and anti--quarks are uncorrelated
[2]. This has allowed to introduce separately the number of light
quarks $q$ and light anti--quarks $\bar{q}$ and the number of strange
quarks $s$ and strange anti--quarks $\bar{s}$ [2]. Moreover, this has
turned out to be of use in order to hide the dependence of quark and
anti--quark distribution functions on a temperature $T$ and a chemical
potential $\mu(T)$ in the number of light quarks $q$ and light
anti--quarks $\bar{q}$. A non--vanishing chemical potential of strange
quarks $\mu_s(T)$ is hidden in the number of strange quarks $s$ and
strange anti--quarks $\bar{s}$. Then, the calculation of
multiplicities of hadrons produced from the QGP phase in a simple
coalescence model resembles a quark counting. In fact, multiplicities
of hadrons are proportional to products of the number of quarks
$(q,s)$ and anti--quarks $(\bar{q}, \bar{s})$ in accord the naive
quark structure of hadrons. Since quarks and anti--quarks do not
correlate, so that the multiplicities of hadrons turn out to be
independent on the momenta of hadrons. Then, the numbers of quarks
$(q,s)$ and anti--quarks $(\bar{q}, \bar{s})$ and the coefficients of
proportionality, the coalescence coefficients $C_{\rm p}$,
$C_{\Lambda}$, $C_{\Xi}$, $C_{\Omega}$ and $C_{\rm \bar{p}}$,
$C_{\bar{\Lambda}}$, $C_{\bar{\Xi}}$, $C_{\bar{\Omega}}$, are free
parameters of a simple coalescence model. Therefore, a total number
free parameters appearing in a simple coalescence model for the
description of baryon and anti--baryon production is equal to
twelve. By the assumption $C_{\rm p}/C_{\rm \bar{p}} =
C_{\Lambda}/C_{\bar{\Lambda}} = C_{\Xi}/C_{\Xi}
=C_{\Omega}/C_{\bar{\Omega}} =1$ the number free parameters has been
reduced up to five $(q,s,\bar{q}, \bar{s}, C)$, where $C$ is a common
for baryons and anti--baryons coalescence coefficient. Two of these
free parameters have been fixed from the fit of experimental data on
the baryon and anti--baryon production: $\bar{q}/q = 0.41\pm 0.02$ and
$\bar{s}/s = 0.75 \pm 0.06$ [2]. Thus, there are three free parameters
left in a simple coalescence model applied to the description of
baryon and anti--baryon production from the QGP phase. It is also
important to note that a simple coalescence model [2] explains only
multiplicities of baryon and anti--baryon production. In fact, save
the ratio of multiplicities of the K$^+$ and K$^-$ mesons none other
multiplicities of pseudoscalar and vector mesons have been predicted
within a simple coalescence model [2]. Therefore, it is not completely
clear how many free parameters would be added in a simple coalescence
model for description of multiplicities of pseudoscalar and vector
meson production.

In our approach, where quarks and anti--quarks coalescing into hadrons
are correlated, we have six free parameters $T$, $\mu(T)$, $C_M$,
$C_B$, $C_{\bar{B}}$ and $F_S$. Five of these parameters $T=175\,{\rm
MeV}$, $\mu(T)$ given by Eq.(\ref{label1.4}), $F_S =3.5\,F_{\pi} =
458.5\,{\rm MeV}$, $C_M/C_B = 0.23\pm 0.04$ and $C_{\bar{B}}/C_B =
0.46\pm 0.15$ are fixed. Therefore, only there is one free parameter
left in the approach. Thus, if to take into account that within our
approach we have described not only multiplicities of baryon and
anti--baryon production from the QGP but also multiplicities of
pseudoscalar and vector meson production, all together 21 experimental
ratios, our approach to the thermalized QGP looks much more preferable
with respect to a simple coalescence model. If to remind that in our
approach due to correlations between quarks and anti--quarks we are
able to follow a dependence of multiplicities of hadron production on
the hadronic momenta, so an advantage of our approach with respect to
a simple coalescence model becomes obvious.

\section*{Acknowledgement}

\hspace{0.2in} One of the authors (A.N. Ivanov) thanks
Prof. T. S. Bir$\acute{\rm o}$, Acad. J. Zim$\acute{\rm a}$nyi and the
members of the Quark--gluon plasma group of Theory Division of
Research Institute of Particle and Nuclear Physics of Hungarian
Academy of Sciences for helpful discussions during the seminar where
this paper has been reported. He is also grateful to
Prof. T. S. Bir$\acute{\rm o}$ and Prof. V. Gogohia for warm
hospitality extended to him during his visit to Budapest.

We are greatful to Prof. A. Rebhan for fruitful
discussions. Discussions with Prof. I. N .Toptygin of the cosmological
aspects of our approach are appreciated.
\newpage

\noindent Table 1. The theoretical ratios of multiplicities of the
hadron production are compared with the experimental data obtained by
NA44, NA49, NA50 and WA97 Collaborations on Pb+Pb collisions at
158\,GeV/nucleon, NA35 Collaboration on S+S collisions and NA38
Collaboration on O+U and S+U collisions at 200\,GeV/nucleon. The
theoretical multiplicities are calculated at the temperature $T =
175\,{\rm MeV}$.
\vspace{0.2in}

\begin{tabular}{|c|c|c|c|c|c|c|}
\hline \cline{2-6}N & Ratio & Model & Data & Coll.& Rapidity & Ref.\\
\hline 1&$\bar{p}/p$ & 0.097(32) & 0.055(10)& NA44 & 2.3--2.9 & [14]\\
& & 0.097(32)& 0.085(8)& NA49 & 2.5--3.3& [15]\\ 2&
$\bar{\Lambda}/\Lambda$& 0.168(55) & 0.128(12)& WA97 &2.4--3.4 &[16]
\\ 3& $\bar{\Xi}/\Xi$& 0.270(88)& 0.227(33)& NA49 &3.1--3.85 &[17] \\
& & 0.270(88)& 0.266(28)& WA97 &2.4--3.4 & [16]\\
4&$\bar{\Omega}/\Omega$& fit & 0.46(15)& WA97 &2.4--3.4 &[16] \\
5&$\Xi/\Lambda$& 0.108 & 0.127(11)& NA49 &3.1--3.85 &[17] \\& & 0.108
& 0.093(7)& WA97 &2.4--3.4 &[16] \\ 6 &$\Omega/\Xi$& 0.166 &
0.195(28)& WA97 &2.4--3.4 &[16] \\ 7&$\bar{\Xi}/\bar{\Lambda}$& 0.173
& 0.180(39)& NA49 &3.1--3.85&[17] \\ & & 0.173 & 0.195(23)& WA97
&2.4--3.4&[16] \\ 8&$\bar{\Lambda}/\bar{p}$& 2.081 & 3(1)& NA49 &
3.1--3.85 &[18] \\ 9&$\bar{\Omega}/\bar{\Xi}$& 0.282 & 0.27(6)& WA97 &
2.4--3.4 &[19] \\ 10&$K^+/K^-$& 1.520 & 1.85(9)& NA44 & 2.4--3.5 &[14]
\\ & & 1.520 & 1.8(1)& NA49 & all &[20] \\ 11&$K^+/\pi^+$& 0.139 &
0.137(8)& NA35 & all &[21] \\ 12&$K^-/\pi^-$& 0.090 &0.076(5) & NA35 &
all &[21] \\ 13 &$K^0_S/\pi^-$& 0.113 & 0.125(19)& NA49 & all &[22] \\
14&$\eta/\pi^0$& 0.088 & 0.081(13)& WA98 & 2.3--2.9 &[23] \\
15&$2\phi/(\pi^+ + \pi^-)$& $7.8\times 10^{-3}$ &$9.1(1.0)\times
10^{-3}$ & NA50 & 2.9--3.9 &[24] \\ 16 &$\phi/(\rho^0 + \omega^0)$&
0.103 &$\approx 0.1$ & NA38 & 2.8--4.1 &[25,26] \\ 17&$\phi/K^0_S$&
0.071& 0.084(11) & NA49 & all &[27] \\ 18&$\Lambda/K^0_S$& fit &
0.65(11)& WA97 & 2.4--3.4 &[28] \\ 19&$p/K^+$& 0.136(23)& & & &\\
20&$\bar{p}/K^-$& 0.020(7)& & & &\\ 21& $\bar{p}/p \cdot K^+/K^-$
&0.147(57) &0.102(19)&NA44 &2.3--2.9 &[14]\\ & &0.147(57)
&0.153(17)&NA49 & 2.5--3.3 &[15,20]\\ \hline
\end{tabular}\\

\end{document}